\documentclass[aps,prd,twocolumn,groupedaddress,showpacs]{revtex4}

\usepackage{graphics}
\usepackage{citesort}
\begin{document}


\title{Binary Black Holes in Quasi-Stationary Circular Orbits}


\author{Brian D.~Baker}
\email[]{bbaker@phys.ufl.edu}
\affiliation{Department of Physics,
             University of Florida,
	     Gainesville, FL 32611}
\date{\today}
\begin{abstract}
We propose a method of determining solutions to the constraint equations of General Relativity approximately describing binary black holes in quasi-stationary circular orbits. Black holes with arbitrary linear momenta are constructed in the manner suggested by Brandt and Br\"{u}gmann. The quasi-stationary circular orbits are determined by local minima in the ADM mass in a manner similar to Baumgarte and Cook; however, rather than fixing the area of the apparent horizon, we fix the value of the bare masses of the holes. We numerically generate an evolutionary sequence of quasi-stationary circular orbits up to and including the innermost stable circular orbit. We compare our results with post-Newtonian expectations as well as the results of Cook and Baumgarte. We also generate additional numerical results describing the dynamics of the geometry due to the emission of gravitational radiation.
\end{abstract}
\pacs{95.30.Sf, 04.70.Bw, 04.20.Ex, 04.25.Dm}
\maketitle
\section{Introduction}
\label{Introduction}
A solution to the fully relativistic two-body problem in General Relativity is the theoretical holy grail of Einstein's theory. The advent of gravitational wave observatories, such as LIGO \cite{Abramovici}, have made the need for a two-body solution even more pressing: without accurate estimates of orbital parameters, investigators have little hope in constructing realistic waveform templates to filter through the avalanche of observational data \cite{FlanaganHughes}.\\
\indent In this paper we present a method of generating solutions to the constraint equations which approximate a binary black hole system in quasi-stationary circular orbits. The quasi-stationary approximation is based on the existence of a helical Killing vector, which implies the system contains equal amounts of inbound and outbound gravitational radiation \cite{Steve1}.\\
\indent We take advantage of the conformally flat conjecture and maximal slicing, which decouples the constraint equations. We also adopt the ``puncture'' method of Brandt and Br\"{u}gmann \cite{BB}, in which our binary system is described by a three-sheeted Brill topology. The puncture method greatly simplifies the analysis of the system, discarding the need for boundary conditions at the throat of the holes. A numerical solution to the Hamiltonian constraint is arrived at via a nonlinear adaptive multigrid algorithm which allows us to impose boundary conditions far from the punctures, while maintaining a relatively high density of grid points in the vicinity of the holes. The quasi-stationary circular orbits are determined via a method based on a formal variational principle which is similar to the ``effective potential'' method of Baumgarte \cite{Baumgarte2} and Cook \cite{Cook4}. The variational principle dictates we must hold the bare masses of the holes fixed, and not the apparent horizon area, as we generate our approximately quasi-stationary configuration. This greatly simplifies numerical implementation, and avoids the need for horizon-finding algorithms.\\
\indent The paper is organized as follows:~Section~\ref{Theoretical Background} is dedicated to the theoretical background pertaining to the initial value problem, including the form of the constraint equations, the puncture method, and the variational principle we employ to generate a sequence of quasi-stationary circular orbits. Section~\ref{Numerical Methods} briefly describes the numerical methods employed, and the tests the code was subjected to. Section~\ref{Numerical Results} presents the numerical results of the code for the binary system in addition to comparing the data to Post-Newtonian expectations and to the results of Baumgarte and Cook. Finally, Section~\ref{Summary} gives a brief summary.
\section{Theoretical Background}
\label{Theoretical Background}
\subsection{The Initial Value Problem}
\label{The Initial Value Problem}
The vacuum constraint equations in the 3+1 formalism \cite{York3+1} are the cornerstone of the initial value problem. The Hamiltonian constraint is given by
\begin{equation}
\label{vp2-1a}
  R + \cal{K}\it^{\;a}_{a} \cal{K}\it^{\;b}_{b} - \cal{K}\it_{ab} \cal{K}\it^{ab} = {\rm 0},
\end{equation}
where $\cal{K}\it_{ab}$ is the extrinsic curvature associated with the three-metric $\gamma_{ab}$, and $R$ is the associated Ricci scalar. The momentum constraint is given by
\begin{equation}
\label{vp2-1b}
  D^{a} \left( \cal{K}\it_{ab} - \cal{K}\it^{\;c}_{c} \gamma_{ab} \right) = 0,
\end{equation}
where $D^{a}$ is the covariant derivative compatible with the three-metric $\gamma_{ab}$. These constraint equations must be satisfied on every spacelike hypersurface, but do not determine the dynamics of the geometry.\\ 
\indent A solution to the initial value problem consists of a three-metric $\gamma_{ab}$ and an extrinsic curvature $\cal{K}\it_{ab}$ which satisfy Eqs.~(\ref{vp2-1a}) and (\ref{vp2-1b}). However, one notes the equations are coupled, nonlinear equations---this greatly complicates any efforts to generate either analytic or numerical solutions to the initial value equations. To simplify the analysis, we assume the three-metric $\gamma_{ab}$ is conformally flat. This introduces the conformal factor $\psi$ via the relationship
\begin{equation}
\label{vp2-3}
 \gamma_{ab} = \psi^{4} g_{ab}, 
\end{equation}
where $g_{ab}$ is the flat three-metric. Following Bowen and York \cite{Bowen}, we have the extrinsic curvature transform according to
\begin{equation}
\label{vp2-4}
 \cal{K}\it_{ab} = \psi^{\rm -2} K_{ab}. 
\end{equation}
We call $K_{ab}$ the conformal extrinsic curvature.\\
\indent In addition to the conformally flat assumption, we choose the maximal slicing gauge \cite{Bowen} which corresponds to the conformal extrinsic curvature being trace-free:
\begin{equation}
\label{vp2-1c}
  K^{\;a}_{a} = 0.
\end{equation}
\indent By assuming a conformally flat three-metric, as well as choosing the maximal slicing gauge, we effectively decouple the constraint equations, making them more amenable to study. In particular, the Hamiltonian constraint equation becomes
\begin{equation}
\label{vp2-5}
  \nabla^{2} \psi + \frac{1}{8} K_{ab} K^{ab} \psi^{-7} = 0,
\end{equation}
where the $\nabla^{2}$ operator is the flat-space Laplacian. Likewise, the momentum constraint simplifies to 
\begin{equation}
\label{vp2-6}
  \nabla_{b} K^{ab}=0. 
\end{equation}
Hence, determination of $K_{ab}$ and $\psi$ via Eq.~(\ref{vp2-5}) and Eq.~(\ref{vp2-6}) completely specify the geometry on an initial space-like hypersurface. In practice, one would take this initial data and evolve it in time via the dynamical Einstein equations.
\subsection{The Puncture Method}
\label{The Puncture Method}
Brandt and Br\"{u}gmann \cite{BB} propose a method which avoids some of the complications introduced by the Misner data set, most notably the requirement of boundary conditions at the throat of the holes \cite{Bowen}. They choose a topology in accordance with the three-sheeted Brill data, and propose a method to compactify the asymptotically flat regions on the lower sheets. This compactification effectively simplifies the domain of integration in which the constraint equations must be solved. Simply stated, with the compactification of an asymptotically flat region to a single point in $R^{3}$, not only is the need to impose boundary conditions on the throat eliminated, but it also eliminates the need to excise domains of integration.\\
\indent Bowen and York \cite{Bowen} note solutions to the momentum constraint, given by Eq.~(\ref{vp2-6}), are well-known for a single black hole with arbitrary linear momentum. The trace-free extrinsic curvature which satisfies the momentum constraint is of the form
\begin{equation}
\label{vp2-7}
  K^{ab}_{(i)} = \frac{3}{2r^2} \left[P^a n^b + P^b n^a - \left(g^{ab} - n^a n^b \right) P^c n_c \right],
\end{equation}
where $P^a$ is the linear momentum of the black hole measured on the upper sheet of the geometry. Because the conformal metric is flat, we may use normal Cartesian coordinates, where $n^a$ is a radial normal vector of the form $n^a = x^a / r$, and $r^2 = x^2 + y^2 + z^2$. For a geometry which consists of $N$ black holes, the conformal extrinsic curvature is given by
\begin{equation}
\label{vp2-8}
  K^{ab} = \sum^{N}_{i=1} K^{ab}_{(i)},
\end{equation}
with each term in the sum having its own coordinate origin. Note for $N$ black holes on the upper sheet, there are a total of $N+1$ sheets in the geometry, each with its own separate asymptotically flat region.\\
\indent Brandt and Br\"{u}gmann then proceed to solve the Hamiltonian constraint, given by Eq.~(\ref{vp2-5}), rewriting the conformal factor $\psi$ as
\begin{equation}
\label{vp2-9}
  \psi = \frac{1}{\alpha} + u,
\end{equation}
where $1/\alpha$ is given by
\begin{equation}
\label{vp2-10}
  \frac {1}{\alpha} = \sum^N _{i=1} \frac{M_{(i)}}{2 \vert \vec{r} - \vec{r}_{(i)} \vert }. 
\end{equation}
The quantity $M_{(i)}$ is the {\em Newtonian} mass of the $i^{\rm th}$ black hole, and $\vec{r}_{(i)}$ is the hole's corresponding location.\\
\indent With these identifications, the Hamiltonian constraint becomes a nonlinear elliptic equation for $u$:
\begin{equation}
\label{vp2-11}
  \nabla^2 u + \beta \left( 1 + \alpha u \right)^{-7} = 0,
\end{equation}
where $ \beta $ is related to the conformal extrinsic curvature $ K^{ab} $ via
\begin{equation}
\label{vp2-12}
  \beta = \frac{1}{8} \alpha^7 K^{ab} K_{ab}.
\end{equation}
\indent Recall we demand asymptotic flatness on the upper sheet, which determines the boundary condition on $u$ at spatial infinity. This Robin boundary condition \cite{Cook3} has
\begin{equation}
\label{vp2-13}
  \frac{\partial u}{\partial r} = \frac{1 - u}{r},
\end{equation}
which demands that $u$ consist of a monopole term as one recedes far from the black holes.\\
\indent Hence, Eq.~(\ref{vp2-7}) coupled to Eq.~(\ref{vp2-11}) and Eq.~(\ref{vp2-13}) complete our mathematical description of the initial value problem.\\
\indent In the framework of the puncture method, it is relatively straightforward to show that the ADM mass \cite{ADM} of a single hole as measured from that hole's asymptotically flat region, called the {\em bare} mass $\overline{m}$, is related to the Newtonian mass $M$ of the hole measured on the upper sheet:
\begin{equation}
\label{vp2-28}
 \overline{m}_{(i)} = M_{(i)} \left( u_{(i)} + \sum_{j \neq i} \frac{ M_{(j)} }{2 D_{ij}} \right),
\end{equation}
where $u_{(i)}$ is the value of the conformal factor correction at the location of the $i^{\text{th}}$ puncture and $D_{ij}$ is the coordinate separation distance between the $i$ and $j$ holes. The total angular momentum $J$ and the puncture separation distance $D_{ij}$ determine each hole's linear momentum $P$ via $J = P D_{ij}$. The values of the linear momentum $P$ and the coordinate separation $D_{ij}$ determine the form of the conformal extrinsic curvature $K_{ab}$ via Eq.~(\ref{vp2-7}).
\subsection{The Variational Principle}
\label{The Variational Principle}
\indent Baker and Detweiler \cite{Det3} formulate a variational principle for binary neutron stars which may be modified for binary black holes in quasi-stationary circular orbits. For fixed values of the bare mass $\overline{m}$ and total angular momentum $J$, solutions to the constraint equations given by Eqs.~(\ref{vp2-6}) and (\ref{vp2-11}) generate an extremum (or stationary point) in the ADM mass only when the geometry is a solution to the quasi-stationary Einstein equations:  
\begin{equation}
\label{stevevp1}
  \delta E_{\text{ADM}} = - 2 N \delta \overline{m} + \Omega \delta J.
\end{equation}
In the above equation, $E_{\text{ADM}}$ is the ADM mass of the system as measured on the upper sheet, $N$ is the ratio of the lapse functions on the lower sheets to that of the upper sheet, $\overline{m}$ is the bare mass of a hole as measured at a lower sheet's asymptotic infinity, $\Omega$ is the orbital angular frequency of the binary system, and $J$ is the total angular momentum of the binary system. Equation~(\ref{stevevp1}) assumes the phases of the inbound and outbound gravitational radiation are held fixed as one generates an evolutionary sequence of quasi-stationary circular orbits.\\ 
\indent The application of the variational principle for binary black holes follows. Specify the value of the total angular momentum $J$ and the bare masses $ \overline{m}$. These quantities are to be held fixed during the analysis. Then, initialize the puncture separation distance $D$.\\
\indent For some initial guess of the Newtonian mass $M$, solve the Hamiltonian constraint equation given by Eq.~(\ref{vp2-11}) subject to the Robin boundary condition given by Eq.~(\ref{vp2-13}). Once $u$ is known, use it to determine new values for the Newtonian masses, as determined by Eq.~(\ref{vp2-28}), so as to hold $\overline{m}$ fixed.\\
\indent Repeat the above procedure by re-solving the Hamiltonian constraint equation using the newly determined Newtonian masses, iterating until the Newtonian masses have converged to within some predetermined tolerance and give the appropriate value of the bare mass. Only then assume that the Hamiltonian constraint is solved subject to fixed values of the total angular momentum $J$ and the bare mass $\overline{m}$. At this point, calculate the ADM mass for the system.\\
\indent Next, decrement the hole separation distance $D$, holding the total angular momentum $J$ and the bare masses $\overline{m}$ fixed to their initial values. Repeat this procedure for a range of hole separations, calculating the ADM mass at each separation distance after reasonable convergence in the Newtonian masses has been obtained.\\
\indent Once the ADM mass for the system has been calculated for a range of separation distances $D$---all the while holding the angular momentum $J$ and the bare masses $\overline{m}$ fixed---we determine the approximate location of a circular orbit via
\begin{equation}
\label{vp2-29}
  \left. \frac{ \partial{ E_{\text{ADM}} } } {\partial {D}} \right\vert_{ \overline{m} ,J }= 0.
\end{equation}
The variational principle also allows us to calculate the orbital angular frequency $\Omega$, given by
\begin{equation}
\label{vp2-30}
 \Omega = \left. \frac{ \partial{ E_{\text{ADM}} } } {\partial {J}} \right\vert_{ \overline{m} }.
\end{equation}
In fact, the variational principle implies that if the initial data are within some small measure $\delta$ of true quasi-stationary data, then $E_{\text{ADM}}$, $J$, $\Omega$, and $N$ are within $\delta^{2}$ of their true quasi-stationary values.\\
\indent Note that our procedure is similar to those of Baumgarte \cite{Baumgarte2} and Cook \cite{Cook4}, but in their heuristic effective potential method they hold the area of the apparent horizons fixed as they generate a member of their evolutionary sequence of circular orbits.
\section{Numerical Methods}
\label{Numerical Methods}
We briefly discuss the numerical methods implemented to solve the nonlinear constraint equation, given by Eq.~(\ref{vp2-11}). We developed an algorithm in C++ to solve the Hamiltonian constraint via adaptive multigrid methods. We chose the multigrid method because of its relative ease of implementation, as well as its speedy convergence rate \cite{numrecipes,multigrid2}. The adaptive portion of the algorithm allows the Robin boundary condition to be imposed arbitrarily far from the holes, subject only to machine memory limitations. This is achieved by starting with a cubical grid comprised of $n^{3}$ grid points, with a physical grid length of $L$. Each successive adaptive grid is also comprised of $n^{3}$ grid points, but has a physical grid length twice that of the previous adaptive grid. This results in a series of embedded grids that yields a high grid density in the area around the holes, while allowing the boundary condition to be imposed far from the holes. Coupled to the symmetry of the equal mass scenario, this allowed for a relatively large number of total grid points to be employed. For instance, if the smallest grid has $n$ grid points on a side and a total of $N$ adaptive levels, the total number of grid points employed in the calculation is on the order of $N n^{3}$. Detailed discussion of the adaptive multigrid algorithm, as well as the complete C++ computer code, may be found in \cite{mythesis}.\\
\subsection{Numerical Tests}
\label{Numerical Tests}
\begin{figure}[]
\begin{center}
\scalebox{0.45}[0.45]{
  \includegraphics{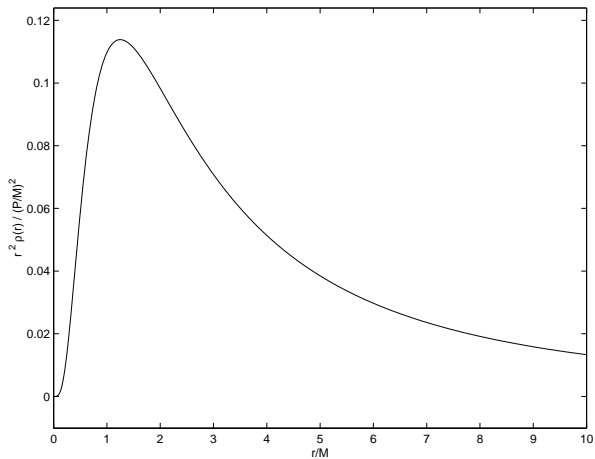}
}
\caption{The radial dependence of the normalized energy density $r^{2} \rho(r) / (P/M)^{2}$ versus the normalized distance $r/M$. The quantity $\rho(r)$ is defined as $1 / 2 \pi \int \beta (1 + \alpha u)^{-7} d\Omega$.}
\label{density3.eps}
\end{center}
\end{figure}

\indent As demonstrated in \cite{BB}, the ADM mass may be determined via
\begin{eqnarray}
\label{test3}
  E_{\text{ADM}} = &&\!\!- \frac{1}{2 \pi} \int_{V} \nabla^{2} \left( \frac{1}{\alpha} \right) d^{3} x \nonumber \\
              &&~~+ \frac{1}{2 \pi} \int_{V} \beta ( 1 + \alpha u )^{-7} d^{3} x.
\end{eqnarray}
\indent It is straightforward to analytically determine the ADM mass for a single boosted black hole in the limit of small $P/M$. Restricting the analysis to $u \approx 1$ and $P/M \ll 1.0$, we find the result to be
\begin{equation}
\label{test7}
  E_{\text{ADM}} = M + \frac{5}{8} \frac{P^{2}}{M}.
\end{equation}
\indent One's attention is immediately drawn to the factor of $5/8$ in front of the ``kinetic'' term. One would prefer the initial value solution described by Eqs.~(\ref{vp2-7})--(\ref{vp2-13}) would yield the expected Newtonian factor of $1/2$. That it does not reflects the fact that even for this simple example, the initial data appears to contain mass-energy in addition to that of the black hole itself. Figure \ref{density3.eps} shows the radial dependence of the energy density; integration of this curve from $r=0$ to $r=\infty$ yields the factor of $5/8$. It is impossible to pinpoint the location of this additional energy; all that can be said is the initial data contains additional energy ``frozen'' into the geometry. Nonetheless, we may use this geometry as a test bed of the code, and we will ultimately apply the variational principle to the geometry to examine binary black holes in quasi-stationary circular orbits.\\
\indent In numerical runs of the single boosted hole with $P/M = 0.1$ and a physical grid length for the smallest grid of $4M$, we found agreement with the theoretical ADM mass correction to be 1.3\%, 0.6\%, and 0.4\% for cubical grids with $n = 11$, $n = 19$, and $n = 35$ grid points on the side of the smallest adaptive grid, respectively. These grid resolutions translate to 2, 4, and 8 grid points across the diameter of the throat of the hole, respectively. For all three resolutions, the Robin boundary condition was imposed at a distance of approximately $4000M$ from the puncture.\\
\indent A second test of the code, as described in \cite{BB}, is two boosted black holes located on an axis at coordinate distances $\pm D$ and moving towards each other. The theoretical expression for the ADM mass is not as straightforward to derive as for a single hole. Once again, assume the correction to the conformal factor $u \approx 1$, and limit the analysis to $P/M \ll 1.0$. After Taylor expanding the second integrand of Eqn.~(\ref{test3}) in factors of $D/M$, we find the ADM mass for the system to be
\begin{eqnarray}
\label{test8}
  E_{\text{ADM}} = 2M + M \left( \frac{P}{M} \right)^{2} \left[ \frac{11}{50} \left( \frac{D}{M} \right)^{2} 
                               - \frac{24}{35} \left( \frac{D}{M} \right)^{4} \right]. \nonumber \\
\end{eqnarray}
Note Eqn.~(\ref{test8}) is accurate only up to fourth order in the expansion parameter $D/M$. In numerical runs with $P/M = 0.1$, $D/M = \pm 0.1$, and a physical grid length for the smallest grid of $4M$, we found agreement with the theoretical ADM mass correction to be 5.4\%, 1.7\%, and 0.9\% for cubical grids with $n = 11$, $n = 19$, and $n = 35$ grid points on the side of the smallest adaptive grid, respectively. Once again, for all three resolutions, the Robin boundary condition was imposed at a distance of approximately $4000M$ from the punctures.
\section{Numerical Results}
\label{Numerical Results}
\begin{figure}[]
\begin{center}
\scalebox{0.45}[0.45]{
  \includegraphics{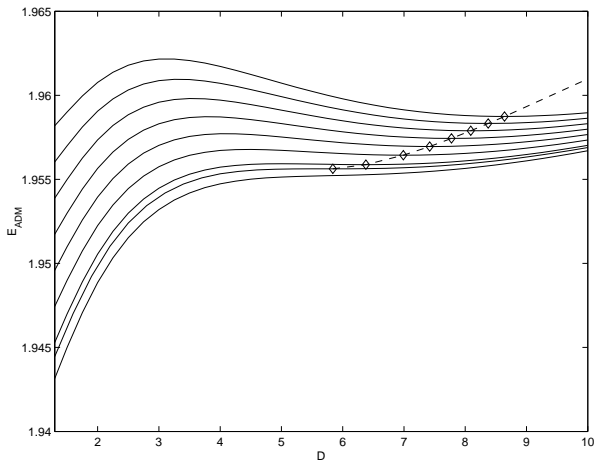}
}
\caption{The ADM mass curves for fixed values of $J$ and $\overline{m}$, and the corresponding evolutionary sequence of circular orbits for $n = 11$ grid points on a side of the smallest adaptive grid. The Robin condition is imposed at a distance of $1000D$. The values of $J$ range from $2.93 {\overline{m}}^{2}$ for the bottommost solid line curve to $3.00 {\overline{m}}^{2}$ for the topmost solid line curve. Our approximation of the evolutionary sequence terminates at $J = 2.937 {\overline{m}}^{2}$, the innermost stable circular orbit.}
\label{MG3curves1.eps}
\end{center}
\end{figure}

\begin{figure}[]
\begin{center}
\scalebox{0.45}[0.45]{
  \includegraphics{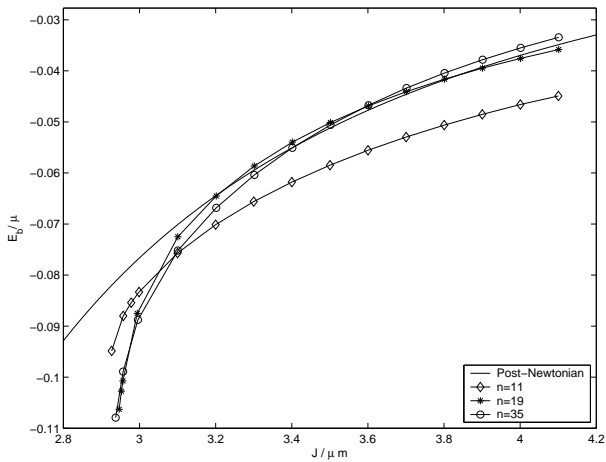}
}
\caption{The normalized binding energy $E_{\text{b}}/ \mu$ versus the normalized angular momentum $J /\mu m$. }
\label{EvJ.eps}
\end{center}
\end{figure}

\begin{figure}[]
\begin{center}
\scalebox{0.45}[0.45]{
  \includegraphics{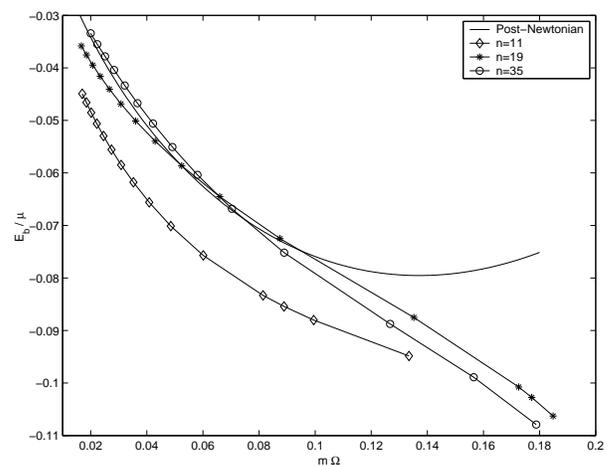}
}
\caption{The normalized binding energy $E_{\text{b}}/ \mu$ versus the normalized angular frequency $m \Omega$. }
\label{EvOmega.eps}
\end{center}
\end{figure}

\begin{figure}[]
\begin{center}
\scalebox{0.45}[0.45]{
  \includegraphics{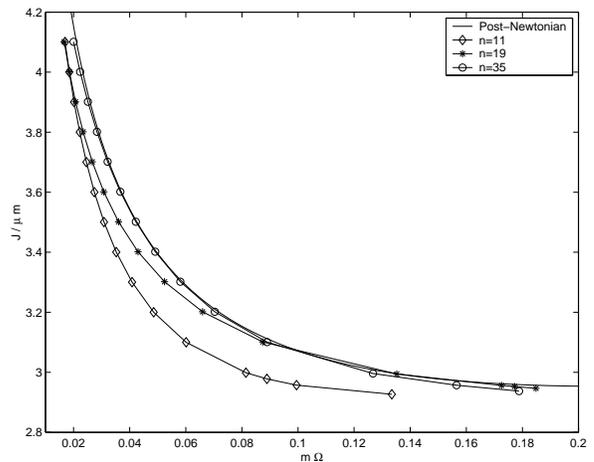}
}
\caption{The normalized angular momentum $J/ \mu m$ versus the normalized angular frequency $m \Omega$. }
\label{JvOmega.eps}
\end{center}
\end{figure}

\indent We now present the numerical results from our adaptive multigrid computer code used to model binary black holes. Only binary systems with black holes having zero intrinsic spin were investigated.\\ 
\indent Figure~\ref{MG3curves1.eps} is an example of the curves generated by the procedure described in Section~\ref{The Variational Principle}, for fixed values of the total angular momentum $J$ and bare mass $\overline{m}$. The vertical axis is the ADM mass of the system, and the horizontal axis is the coordinate separation distance between the two holes. The location of the circular orbits is denoted by the dashed curve, with diamonds representing the locations of the circular orbits for particular values of $J$. This particular figure was generated for a grid with $n = 11$ grid points on the side of the smallest adaptive grid, with the Robin boundary condition imposed at approximately $1000D$, where $D$ is the coordinate separation distance of the holes. The bottommost solid line curve corresponds to a value of the angular momentum $J = 2.93 \overline{m}^{2}$, and the topmost solid line curve corresponds to a value of $J = 3.00 \overline{m}^{2}$. The evolutionary sequence of circular orbits terminates at a coordinate separation distance of $D=5.8395 \overline{m}$ and total angular momentum $J = 2.937 \overline{m}^{2}$, which corresponds to the innermost stable circular orbit.\\
\indent The resulting evolutionary sequence of circular orbits, denoted by the dashed line, can be viewed in the following fashion. A realistic binary system will radiate gravitational waves, causing their orbital separation distance to decrease slowly. The binary system will progress along the evolutionary sequence, until the binary system reaches the innermost stable circular orbit. It is after the innermost stable circular orbit that we may no longer employ our variational principle, simply due to the lack of stable circular orbits for smaller separation distances. Because of this, we can not determine the dynamics of the final inward spiral and plunge. Despite this shortcoming, the variational principle, and the evolutionary sequence it generates, yields valuable information about the evolution of the system up to and including the innermost stable circular orbit.\\
\indent One problem that arises in the analysis is the question of black hole mass. Specifically, what does one mean when one speaks of the mass of a black hole in a binary system? There is no general consensus on the answer to this question. Recall from Section~\ref{The Puncture Method} that we encountered two types of mass when describing the punctures of Brandt and Br\"{u}gmann: the bare mass $\overline{m}$ and the Newtonian mass $M$.\\
\indent However, these masses are not the only used to describe black holes. For instance, Cook \cite{Cook4} and Baumgarte \cite{Baumgarte2} choose to use the mass associated with the area of the apparent horizon, described by Christodoulou's formula \cite{Christo}.\\
\indent A different measure of the mass of a black hole is the rest mass \cite{Cook3}, described by the well-known formula from special relativity:
\begin{equation}
\label{numresults2}
  E_{\text{rest}} = \sqrt{ E_{\text{ADM}}^{2} - P^{2} },
\end{equation}
where $E_{\text{ADM}}$ is the ADM mass of an isolated black hole with linear momentum $P$. We have opted to use the rest mass when comparing our numerical results to the expected post-Newtonian results, simply because its value depends only upon the properties of the single hole, and not the interaction energy between the holes. However, $E_{\text{rest}}$ includes the additional energy which is ``frozen'' into the initial data, and would be expected to effect the dynamics of the system. For a binary system with each hole having a linear momentum $P$, the rest mass of each hole is calculated by numerically evaluating the ADM mass for a single, isolated, boosted black hole with linear momentum $P$.\\
\indent Kidder et al. \cite{Kidder} provide $($post$)^{2}$-Newtonian expressions for the total angular momentum $J$, orbital angular frequency $\Omega$, and the binding energy $E_{\text{b}}$. Instead of calculating the binding energy based on the mass associated with the apparent horizon \cite{Cook4,Baumgarte2}, we define the binding energy to be
\begin{equation}
\label{numresults7}
  E_{\text{b}} \equiv E_{\text{ADM}} - 2 E_{\text{rest}}.
\end{equation}
We also introduce the reduced mass $\mu$ and the total mass $m$, defined via
\begin{equation}
\label{numresults6}
  \mu \equiv \frac{1}{2} E_{\text{rest}}
\end{equation}
and
\begin{equation}
\label{numresults5}
  m \equiv 2 E_{\text{rest}}.
\end{equation}
\indent The $($post$)^{2}$-Newtonian equations, written in the form of Cook \cite{Cook4}, are
\begin{eqnarray}
\label{pn1}
  \frac{E_{\text{b}}}{\mu} = - \frac{1}{2} \left( \frac{\mu m}{J} \right)^{2}&& \!\!\!\!\!\!
                       \left[ 1 + \frac{37}{16} \left( \frac{\mu m}{J} \right)^{2} \right. \nonumber \\
		       &&~~\left. + \frac{1269}{128} \left( \frac{\mu m}{J} \right)^{4} + \ldots \right],
\end{eqnarray}

\begin{eqnarray}
\label{pn2}
  \frac{E_{\text{b}}}{\mu} = - \frac{1}{2} (m \Omega)^{2/3}&& \!\!\!\!\!\!
                       \left[ 1 - \frac{37}{48} (m \Omega)^{2/3} \right. \nonumber \\
		       &&~~\left. - \frac{1069}{384} (m \Omega)^{4/3} + \ldots \right],
\end{eqnarray}
and finally
\begin{eqnarray}
\label{pn3}
  \left( \frac{J}{\mu m} \right)^{2} = (m \Omega)^{-2/3}&& \!\!\!\!\!\!
                                        \left[ 1 + \frac{37}{12} (m \Omega)^{2/3} \right. \nonumber \\
					&&~~\left. + \frac{143}{18} (m \Omega)^{4/3} + \ldots  \right].
\end{eqnarray}
\indent We now present the figures of our numerical results, compared to $($post$)^{2}$-Newtonian expectations.\\
\indent Figure~\ref{EvJ.eps} displays the normalized binding energy $E_{\text{b}}/\mu$ versus the normalized angular momentum $J/\mu m$. The solid line is the theoretical prediction given by Eq.~(\ref{pn1}). One notes that the curves corresponding to $n=19$ and $n=35$ actually cross the theoretical prediction, as opposed to asymptotically approaching the solid line. We believe that if the grid resolution were to be increased, the resulting data would conform more closely to the post-Newtonian prediction for large $J/\mu m$.\\
\indent Figure~\ref{EvOmega.eps} displays the normalized binding energy $E_{\text{b}}/\mu$ versus the normalized angular frequency $m \Omega$. The solid line is the theoretical prediction given by Eq.~(\ref{pn2}). Note the Newtonian regime of the numerical curves, corresponding to smaller values of $m \Omega$, agree with the post-Newtonian prediction. There is agreement between the theoretical prediction and the two higher resolution curves up until $m \Omega \approx 0.08$. At this point, the curvature of the two higher resolution curves deviates from the theoretical prediction, and tends to more negative values of the normalized binding energy.\\
\indent Figure~\ref{JvOmega.eps} displays the normalized angular momentum $J/\mu m$ versus the normalized angular frequency $m \Omega$. The theoretical prediction is given by Eq.~(\ref{pn3}), and is denoted in the figure by the solid line. The curves indicate convergence at both the Newtonian and the relativistic ends of the figure.\\
\indent Before we compare the data to that of Baumgarte \cite{Baumgarte2} and Cook \cite{Cook4}, we feel it is important once again to stress that there is some ambiguity as to what one means when one talks about the mass of a black hole in a binary system. Both Cook and Baumgarte associate the area of the apparent horizon to the mass of an individual hole, and hold the area of the apparent horizon fixed as they generate their evolutionary sequence of circular orbits. We, on the other hand, choose to use the rest mass to describe the individual hole, and held the bare mass of each hole fixed as we generated our evolutionary sequence of circular orbits.\\
\indent We begin by noting that we cannot compare our coordinate separation distance $D/m$ to the proper separation distances determined by Baumgarte and Cook. Our main limitation in this is our lack of knowledge concerning the locations of the apparent horizons of the holes. However, we may still get an approximate measure $d$ of the separation distance of the holes at the innermost stable circular orbit: we assume the quadrupole moment of the system, as measured at infinity, consists of two objects of mass $\frac{1}{2} E_{\text{ADM}}$ separated by the distance $d$. We present the innermost stable circular orbit's normalized coordinate separation $D/m$, as well as the separation distance $d$ normalized by the system's ADM mass, in Table \ref{quaddisttable}.\\
\renewcommand{\arraystretch}{1.5}
\begin{table}[]
\begin{center}
\caption{The normalized coordinate separation distance $D/m$, and the normalized separation distance $d/E_{\text{ADM}}$ for three resolutions at the innermost stable circular orbit.}
\label{quaddisttable}
\begin{tabular}{| c | c | c |}
\multicolumn{3}{}{} \\  
  \hline Resolution  & $D/m$ & $d/E_{\text{ADM}}$ \\
  \hline n = 11      & 2.915 & 3.078\\
  \hline n = 19      & 2.143 & 2.290\\
  \hline n = 35      & 2.195 & 2.349\\ \hline
\end{tabular} 
\end{center}
\end{table}
\indent Table \ref{datacompare} compares the results for the innermost stable circular orbit for the highest resolution of this work ($n = 35$ grid points on the side of the smallest adaptive grid) to that of Cook \cite{Cook4} and Baumgarte \cite{Baumgarte2}, as listed in Baumgarte.\\
\renewcommand{\arraystretch}{1.5}
\begin{table}[]
\begin{center}
\caption{Comparison of the innermost stable circular orbit data for Cook, Baumgarte, and this work.}
\label{datacompare}
\begin{tabular}{| c | c | c | c |}
\multicolumn{4}{}{} \\  
  \hline Data & $E_{\text{b}}/\mu$ & $J/\mu m$ & $m \Omega$ \\
  \hline Cook      & -0.09030 & 2.976 & 0.172\\
  \hline Baumgarte & -0.092   & 2.95  & 0.18\\
  \hline This work & -0.10792 & 2.937 & 0.179\\ \hline
\end{tabular} 
\end{center}
\end{table}
\indent The agreement between our data and that of Cook and Baumgarte may lend some credence both to their method of approach, as well to the variational principle we employed. On one hand, our method is based on a mathematically derived variational principle for binary black holes in quasi-stationary circular orbits in which the variational principle dictates we must hold the bare mass fixed, and not the area of the apparent horizon. This greatly simplifies the analysis as one does not need to be concerned with the location of the apparent horizon, which can be a daunting numerical task \cite{Cook2,Baumgarte1}. On the other hand, the agreement of Cook and Baumgarte's results with ours may indicate that there is some interesting physics yet to be discovered concerning the relationship between the variational principle we employed and the methods employed by Cook and Baumgarte.
\subsection{Gravitational Waves}
\label{Gravitational Waves}
\begin{figure}[]
\begin{center}
\scalebox{0.45}[0.45]{
  \includegraphics{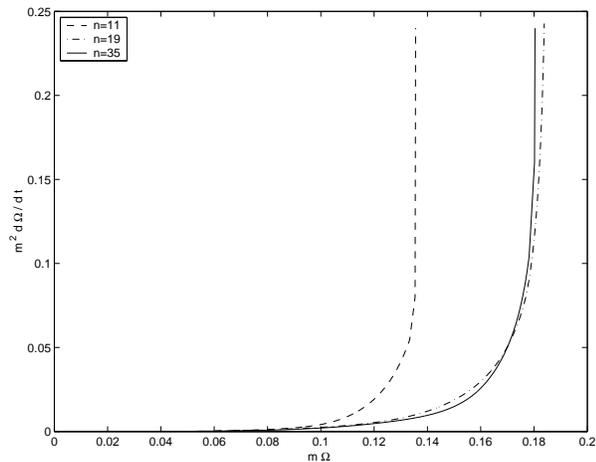}
}
\caption{The normalized rate of change of the orbital angular frequency $m^{2} d\Omega/dt$ versus the normalized orbital angular frequency $m \Omega$. As the system approaches the innermost stable circular orbit, the orbit evolves so rapidly that $m^{2} d\Omega/dt$ effectively diverges, as indicated by the nearly vertical asymptotes.}
\label{newdOmegadt.eps}
\end{center}
\end{figure}

\begin{figure}[]
\begin{center}
\scalebox{0.45}[0.45]{
  \includegraphics{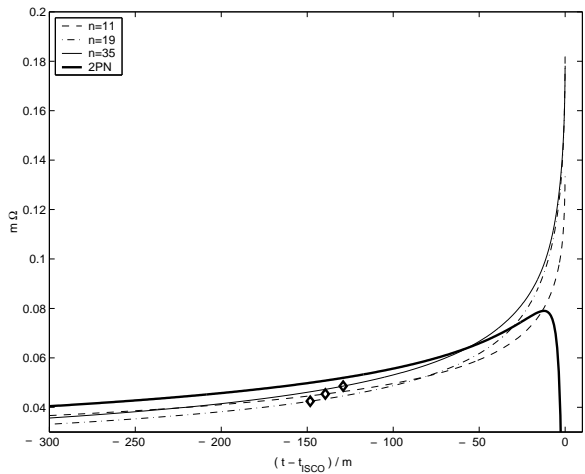}
}
\caption{The normalized orbital angular frequency $m \Omega$ versus the normalized time scale $\left(t-t_{\text{ISCO}}\right)/m$. For a particular data set, the diamond on each curve marks $t_{\text{BLO}}$, the approximate time for one more complete orbit before reaching the innermost stable circular orbit.}
\label{OmegaTnormal.eps}
\end{center}
\end{figure}

\indent An algorithm was developed to calculate the quadrupole moment of the field, which yields the dimensionless luminosity of the gravitational waves via the quadrupole moment formula \cite{MTW}, and subsequently determines the dynamics of our binary system.\\
\begin{figure}[]
\begin{center}
\scalebox{0.45}[0.45]{
  \includegraphics{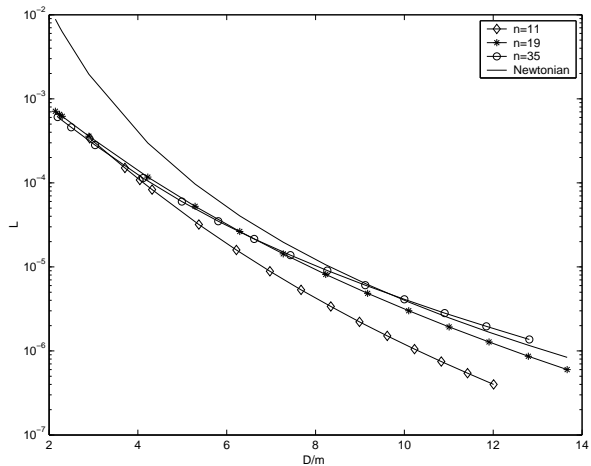}
}
\caption{The dimensionless gravitational wave luminosity $L$ as a function of the normalized separation distance $D/m$.}
\label{LvD.eps}
\end{center}
\end{figure}
\indent Figure~\ref{LvD.eps} displays the gravitational wave luminosity $L$ as a function of the normalized separation distance $D/m$. The three different resolutions are displayed, as well as the Newtonian expectation for the gravitational wave luminosity.  The numerical results appear to be converging to a limit which is in line with the Newtonian expectation for large separation distances. Also, the numerical results are in excellent agreement with each other as the separation distance approaches the innermost stable circular orbit.\\
\indent The quasi-stationary approximation is a powerful tool, but it is natural to inquire about its range of validity. Specifically, can the information extracted from the gravitational radiation indicate a point in which we may no longer rely upon the quasi-stationary approximation to describe our binary system?\\
\indent We begin by noting the gravitational wave luminosity $L$ may be used to determine the time derivative of the orbital angular frequency $\Omega$ via
\begin{equation}
\label{break1}
  \frac{d\Omega}{dt} = \left( \frac{dE_{\text{ADM}}}{d\Omega} \right)^{-1} L.
\end{equation}
The results of this calculation for the three resolutions under study are shown in Fig.~\ref{newdOmegadt.eps}, which shows $m^2 d\Omega/dt$ versus the normalized orbital angular frequency $m \Omega$. As indicated in the figure, as the system approaches the innermost stable circular orbit, the rate of change of the angular frequency diverges.\\
\indent The information from $d\Omega/dt$ may in turn be used to yield the orbital angular frequency as a function of time. This is achieved via
\begin{equation}
\label{break2}
  t - t_{\text{ISCO}} = \int^{\Omega}_{\Omega_{\text{ISCO}}} \frac{d\Omega}{\left( \frac{d\Omega}{dt} \right)}.
\end{equation}
The quantity $t - t_{\text{ISCO}}$ is a negative number, and is a measure of the amount of time that must elapse before the system reaches the innermost stable circular orbit. Figure~\ref{OmegaTnormal.eps} shows the normalized orbital angular frequency $m \Omega$ versus the normalized time scale $\left(t - t_{\text{ISCO}}\right)/m$. The solid bold curve is the $($post$)^{2}$-Newtonian equation of Blanchet et al. \cite{BlanchetDamour}. The diamond on each numerical curve corresponds to the time when the system begins its last orbit, $t_{\text{BLO}}$, before reaching the innermost stable circular orbit; i.e.,
\begin{equation}
\label{break3}
  \frac{\Omega \vert t_{\text{BLO}}-t_{\text{ISCO}}\vert}{2\pi} = 1
\end{equation}
is satisfied.\\
\indent In Table \ref{breaktable}, we list various parameters associated with the beginning of the last orbit, or ``BLO'', as determined by Eq.~(\ref{break3}). We compare this with the values of the innermost stable circular orbit, or ``ISCO'', as determined by the evolutionary sequence of quasi-stationary circular orbits. All values are taken from the highest resolution data. Table \ref{breaktable} shows the parameters of the BLO are far removed from the parameters associated with the ISCO. However, the value of $v/c \approx 0.25$ for the BLO indicates the quasi-stationary method probes regimes where post-Newtonian approximations may start to break down.
\renewcommand{\arraystretch}{1.5}
\begin{table}[]
\begin{center}
\caption{Comparisons between the quasi-stationary ISCO and the beginning of the last orbit, or BLO.}
\label{breaktable}
\begin{tabular}{| c | c | c | c | c |}
\multicolumn{5}{}{} \\  
  \hline \mbox{}& $m \Omega $ & $J/\mu m$  & $D/m$ & $v/c$ \\
  \hline ISCO & 0.179   & 2.937&  2.19 & 0.56\\
  \hline BLO  & 0.049   & 3.401  &  6.62 & 0.25\\ \hline
\end{tabular} 
\end{center}
\end{table}
\section{Summary}
\label{Summary}
\indent In this paper, we propose a method of generating sequences of approximate quasi-stationary circular orbits for binary black hole systems similar to the method of Cook \cite{Cook4} and Baumgarte \cite{Baumgarte2}. The method is based on the variational principle of Baker and Detweiler \cite{Det3}, and employs the ``puncture'' method of Brandt and Br\"{u}gmann \cite{BB}.\\
\indent We developed an adaptive multigrid algorithm which solved the nonlinear Hamiltonian constraint and allowed us to extract useful orbital parameters for the system, up to and including the innermost stable circular orbit. Despite the additional energy content of the geometry, our results agree with post-Newtonian expectations, and are in line with the work of Baumgarte \cite{Baumgarte2} and Cook \cite{Cook4}. We also determined, via extraction of gravitational wave information, the beginning of the last orbit, or BLO.\\
\indent Our method is straightforward as holding the bare mass fixed does not require locating the apparent horizon of the holes. Coupled to the puncture method of Brandt and Br\"{u}gmann, numerical implementation via adaptive multigrid methods is relatively easy.\\
\indent This paper reveals the power of the variational principle. Relaxation of the conformally flat conjecture is the logical next step. This will couple the constraint equations and introduce complications in determination of a solution, but it will eliminate the additional energy content of the geometry, and allow for more realistic simulations of quasi-stationary binary black holes. 

\begin{acknowledgments}
It is a pleasure to thank Steven Detweiler for support, guidance, and insight during this project. The author would also like to thank Eirini Messaritaki for careful reading of the manuscript and useful comments.
\end{acknowledgments}

\end{document}